# DRC-TTS: A Dual-Resolution Codec-Based Architecture for Lexicon-Free Malay Speech Synthesis

**DRC-TTS: A Dual-Resolution Codec-Based Architecture for Lexicon-Free Malay Speech Synthesis**

**Tianhui Su, Tien-Ping Tan***

School of Computer Sciences, Universiti Sains Malaysia, 11800 USM, Penang, Malaysia

*E-mail addresses* tienping@usm.my

*Corresponding author

List of Table/Figure: Table 1. Table 2. Table 3. Table 4. Table 5. Table 6. Table 7. Figure 1. Figure 2. Figure 3. Figure 4. Figure 5. Figure 6. Figure 7.

**Title:** DRC-TTS: A Dual-Resolution Codec-Based Architecture for Lexicon-Free Malay Speech Synthesis

**Abstract:**

Modern discrete neural audio codecs facilitate speech synthesis by replacing continuous acoustic representations with low-frame-rate discrete token sequences. For Malay speech synthesis, character-level input reduces reliance on pronunciation lexicons. However, combining character-level inputs with low-frame-rate discrete codecs introduces a temporal alignment challenge. When predicted character durations are quantized into frame counts at 12.5 Hz, durations shorter than one codec frame may be assigned zero frames. Consequently, the corresponding character representations may be lost during length regulation, potentially leading to pronunciation omissions. This study proposes DRC-TTS, a dual-resolution codec-based architecture for lexicon-free Malay speech synthesis. The model expands character representations at a temporal resolution of 50 Hz and subsequently downsamples them through strided convolution to 12.5 Hz for discrete codec-token prediction. Objective evaluations show that the proposed model reduces the character error rate (CER) from 27.11% for the baseline to 21.86%, an absolute reduction of 5.25 percentage points. The proposed model also achieves a mean opinion score (MOS) of 4.30, compared with 4.12 for the baseline. These results suggest that the dual-resolution design may improve character-level alignment and perceived speech naturalness under the evaluated conditions.

## INTRODUCTION

The advancement of deep learning architectures has accelerated the advancement of text-to-speech (TTS) synthesis, transitioning from conventional TTS methods to end-to-end neural networks that generate intelligible artificial voices [1], [2], [3]. The deployment of synthesis systems expands digital accessibility and automated human computer interactions [4], [5]. A recent paradigm shift in this domain involves the adoption of discrete neural audio codecs in TTS to allow speech to stream over the network at low bandwidth [6], [7]. These codecs compress continuous waveforms into discrete acoustic tokens to eliminate the computational requirement associated with continuous vocoders [8]. The neural codecs operate at low acoustic frame rates to maximize compression efficiency [9]. Figure 1 summarizes how the text frontend, duration-based acoustic model, and neural codec can be combined. The neural TTS trains and learns to predict the codec tokens instead of the speech spectrum. The codec tokens transmit through the network and arrive at the target machine. The audio codec decoder then decodes the codec tokens back to speech. During training, reference audio supplies codec-token targets through codec's encoder and quantizer, while forced alignment supplies text/phonemic transcription-duration targets that is important to produce natural speech. During inference, the model predicts discrete codec tokens from the text/phonemic transcription, and the pretrained codec decoder converts those tokens into speech. In duration based non autoregressive systems, parallel generation requires an explicit mapping between input tokens and acoustic frames [10]. The temporal resolution of these frames therefore affects how short duration input tokens are represented.

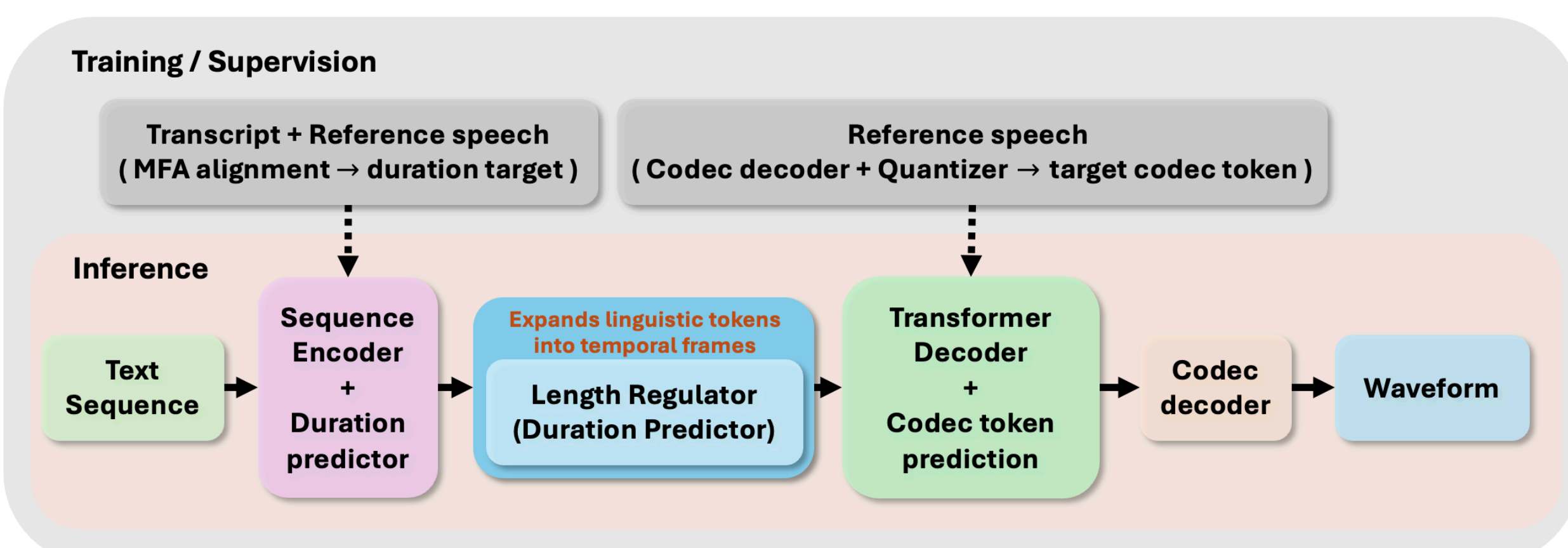

*Figure 1.* Overview of TTS and codec pipeline

The Malay language is a low resource language in speech processing. There are some Malay text to speech systems, but there are not much study on Malay TTS with neural codec. In this study, we investigate how non-autoregressive TTS interacts with low-rate discrete codec tokens, focusing on a temporal alignment problem in duration-based synthesis. Unlike traditional models that generate speech piece by piece, a non-autoregressive system generates the entire audio sequence at the same time. To achieve this, the system must first predict exactly how long each text/phonemic will sound and stretch the text to match that physical time. In addition, Existing Malay synthesis models frequently rely on pronunciation lexicons to map words to phonetic representations. Because Malay features a relatively regular letter to phoneme mapping, utilizing character-level modeling as input tokens mitigates this reliance on explicit dictionaries [11]. However, removing the pronunciation dictionary introduces the letter to codec token alignment during training and decoding because more one or more letters may align to a single codec token.

Figure 2 illustrates the word "SATU" and the duration for each character. Each character has to be aligned to a one or more codec token during training, so that the decoder knows the token to generate given the characters. However, a short character such as "T" in this case has to be handled properly as it may either generate no token at all, or both "T" and "U" may generate the same token. As shown in the visual example, a 12.5 Hz system groups audio into 80 millisecond frames. If a short consonant lasts for less than 40 milliseconds, if a rounding rule is used, no token will be assigned to the character "T" and the decoder will have a problem during training.

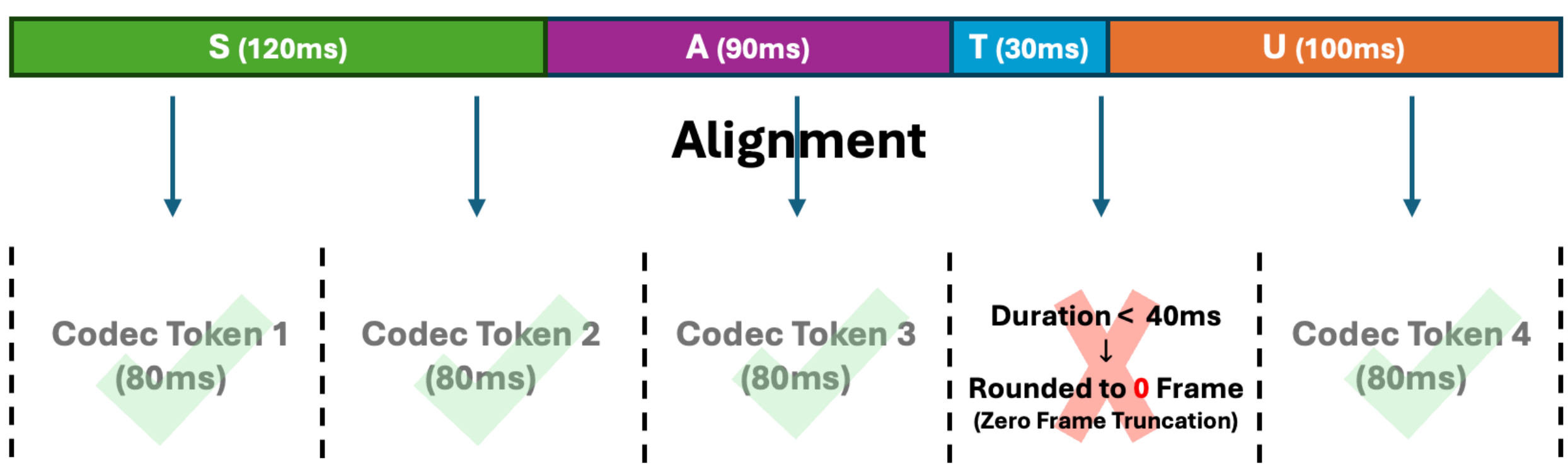

*Figure 2.* Illustration of the zero frame truncation process

The research contributions of this paper are two folds. First, it investigates the integration of Malay text-to-speech (TTS) synthesis with a discrete neural audio codec, focusing on the temporal alignment challenges introduced by low-frame-rate token generation. To overcome the truncation of short consonants caused by low framerate codec quantization, this study proposes a dual-resolution architecture that expands character representations at a higher temporal resolution of 50 Hz before downsampling them to 12.5 Hz for discrete codec-token prediction. This design preserves intermediate temporal representations while maintaining compatibility with the codec's lower frame rate, thereby reducing zero-frame quantization without increasing the number of codec frames generated.

Second, the research introduces a synthesis pipeline that trains the model to predict audio codec tokens directly from letters, which does not require a pronunciation dictionary. A pronunciation dictionary is used to convert letters/characters to phoneme that better matches speech sounds. For example, letter 'e' can be pronounced as /ə/ as in "emak" or /e/ as in "elok". However, pronunciation dictionary often requires human experts to prepare. Even though some automatic methods exist, wrong prediction of the pronunciation can produce cascading effect in a speech synthesis system. Nevertheless, removing the pronunciation dictionary from the speech synthesis pipeline creates other problems. One of the problems is the problem of aligning character to codec token. By mapping character to discrete codec tokens, this pronunciation dictionary free method simplifies the data processing workflow and provides a practical design for low resource languages.

## LITERATURE REVIEW

### Malay Speech Synthesis and Resource Constraints

The advancement of high-quality Malay text-to-speech (TTS) systems is constrained by the availability, scale, and annotation quality of speech corpora. Recent studies from the last few years have introduced several open-source Malay speech datasets, such as the Malaya-Speech corpus and conversational speech datasets provided by MagicHub [12], [13]. Existing resources differ considerably in recording conditions, speaker coverage, linguistic annotations, and intended applications. For example, the IIUM subset of Malaya-Speech contains approximately 2.4 hours of recorded speech from two speakers, illustrating the limited scale of some publicly documented Malay speech resources [13]. The Syntactic-Prosody of Malay Speech Corpus provides complementary linguistic information, comprising 422 sentences, 1,720 phrases, and 6,978 word units linked to syntactic–prosodic structures [14]. More recent resources also include large-scale synthetic speech. Malaysian-TTS-v2 reports approximately 4,642.77 hours of synthetic audio from two speakers [15]. In contrast, the Malay Multi-speaker Speech Synthesis Corpus (Multi-emotion) provides approximately 10.84 hours of professionally recorded speech from three speakers, covering multiple content domains and emotional styles [16]. These resources represent different approaches to addressing Malay TTS data requirements: small-scale recorded speech, linguistically annotated corpora, large-scale synthetic data, and multi-speaker recordings with broader expressive coverage. Their differences also indicate that corpus size alone is insufficient to determine suitability for TTS training; recording quality, speaker diversity, text–speech consistency, annotation reliability, and data accessibility must also be considered. Table 1 summarizes the principal acoustic and linguistic considerations for preparing TTS corpora.

Table 1

*Key Acoustic and Linguistic Requirements for TTS Corpora*

| Feature | Typical TTS Corpora |
|---|---|
| **Acoustic Environment** | Clear recordings with limited background noise and minimal recording artifacts. |
| **Speaker Characteristics** | Consistent speaker identity and adequately documented speaker attributes. |
| **Text Normalization** | Consistent normalization of numbers, abbreviations, symbols, and other written forms. |
| **Text-Speech Alignment** | Accurate correspondence between the input text and the associated speech recordings |

Beyond acoustic properties, existing Malay resources often lack standardized text normalization protocols [17]. Natural Malay text contains complex semiotic classes, including localized currency formats and numerical abbreviations, which require explicit expansion into readable sequences [18]. In terms of text preprocessing, other existing Malay TTS studies frequently rely on grapheme to phoneme conversion pipelines adapted from related languages or rule-based phonetic dictionaries [19], [20]. For instance, previous research has utilized Indonesian phonemicization tools or statistical models to estimate Malay pronunciations prior to acoustic training [21] . These conventional preprocessing pipelines map expanded words to phonetic representations but depend heavily on pronunciation lexicons. Lexicon-based approaches may fail to provide accurate mappings when encountering loanwords or domain-specific terminology [4]. Addressing these vocabulary limitations without relying on continuous manual lexicon updates requires an alternative text representation strategy.

### Text Normalization and Character Level Modeling

Modern neural models traditionally require pronunciation lexicons to generalize accurately across diverse vocabularies [20]. For low resource languages, constructing these dictionaries is a labor intensive process [22]. Cross lingual transfer learning often applies established prediction rules from related languages to circumvent this limitation [23]. However, standard Malay exhibits distinct orthographic conventions compared to linguistic siblings like Indonesian [19]. External phonetic architectures therefore produce alignment errors during acoustic model training.

To mitigate these incompatibilities, some studies advocate for character level modeling in languages with orthographic transparency. Malay exhibits a shallow orthography where written characters correlate consistently with spoken acoustic units. Utilizing characters directly as the input sequence reduces alignment errors caused by out of vocabulary terms [2]. Existing sequence encoders process these character inputs to learn context dependent representations implicitly [24]. Coupling this tokenization with a text normalization pipeline ensures that numerical expressions and localized abbreviations are expanded into readable formats [18], [25]. This combined methodology provides the deterministic sequence necessary for modern forced alignment tools without relying on explicit phonetic dictionaries [26].

While character level modeling removes the dependence on pronunciation lexicons, it introduces a temporal challenge in duration based architectures. Alphabetical characters correspond to brief acoustic events. The physical duration of individual consonants frequently lasts less than 40 milliseconds [27]. Integrating character tokens with low frame rate discrete codecs creates an alignment limitation. In duration based non autoregressive systems, when character duration estimates are rounded to frame counts at a low acoustic resolution, short assigned durations receive zero frames [1]. This duration rounding leads to pronunciation omissions. The specific alignment conflict between character level inputs and low frame rate discrete codecs remains unresolved, requiring a structural modification to preserve short tokens during acoustic generation.

### Advancements in Duration Based Neural TTS Architectures

The landscape of neural speech synthesis shifted from autoregressive generation to parallel non autoregressive architectures. Early models generated speech sequentially frame by frame, resulting in slow inference speeds [28]. To achieve parallel generation, the field introduced explicit duration modeling. Foundational models including FastSpeech [10] and FastSpeech 2 [1] utilize a variance adaptor containing a duration predictor to estimate the number of acoustic frames each input token should occupy. A length regulator subsequently duplicates the hidden states of the input tokens according to these predicted durations.

Subsequent advancements expanded this non autoregressive paradigm to improve alignment precision and prosodic modeling. Architectures such as Glow TTS [29] and VITS [30] employed monotonic alignment search and stochastic duration predictors to model temporal variations dynamically. Models including Parallel Tacotron [2] and PortaSpeech [31] integrated differentiable duration modeling and normalizing flows to refine length regulation. More recent frameworks including StyleTTS [32] and NaturalSpeech [33] inherited these duration prediction mechanisms to drive latent speech generation. Despite utilizing diverse generative decoders ranging from adversarial networks to flow based models, these parallel synthesis architectures remain duration based. They rely on length regulation mechanisms to duplicate the hidden states of linguistic tokens to match the target acoustic sequence length.

This length regulation mechanism requires a deterministic mapping between the linguistic tokens and the acoustic frames. The temporal resolution of the acoustic frames defines the minimum assignable duration for any token. If the acoustic frame rate is high, the length regulator can assign frame counts to short text units. However, reducing the frame rate increases the physical duration of a single frame. When character duration estimates are rounded to frame counts at a low acoustic resolution, short assigned durations may round to zero. This duration rounding leaves duration based non autoregressive models vulnerable to pronunciation omissions when processing extremely short text units.

### Discrete Audio Codecs and Temporal Alignment Constraints

A development in speech synthesis involves replacing continuous acoustic features with discrete neural audio codecs. Inspired by advancements in language modeling, architectures such as VALL-E [6] and AudioLM [34] utilize discrete codecs to compress audio into compact token sequences. Modern discrete codecs, including SoundStream [9], EnCodec [8], Descript Audio Codec [35], and Mimi [7], eliminate the computational overhead associated with traditional continuous vocoders. These neural codecs consist of an encoder network, a residual vector quantizer, and a decoder network. The encoder projects the continuous waveform into a latent space, while the quantizer discretizes these representations into multiple hierarchical codebooks.

To maximize compression efficiency and reduce the sequence length for the generation model, neural codecs employ large strided convolutions within their encoder networks. This structural design produces extremely low acoustic frame rates [8], [9]. For example, a codec operating at 12.5 Hz generates one acoustic frame every 80 milliseconds. While this low resolution accelerates the discrete token generation process, it imposes strict limitations on the representation of rapid acoustic events. Linguistic features that occur within a temporal window shorter than a single acoustic frame are compressed into a singular discrete representation, masking the micro acoustic boundaries required for temporal modeling.

Investigating this codec integration in the context of character level Malay synthesis reveals a temporal alignment constraint. As established in the preceding sections, character tokens frequently exhibit physical durations shorter than 40 milliseconds. In duration based non autoregressive systems, when these duration estimates are rounded to frame counts at the 12.5 Hz codec resolution, the sub threshold durations receive zero frames. These tokens therefore contribute no expanded hidden states during length regulation. Current discrete synthesis models primarily employ autoregressive approaches that predict tokens sequentially, avoiding explicit duration assignments [6], [34]. Existing parallel discrete models have not extensively investigated the integration of character level inputs under low resource conditions. The specific alignment conflict between short character level inputs and low frame rate discrete codecs in parallel architectures remains unresolved. This unresolved resolution conflict requires an architectural modification to preserve short tokens without changing the efficiency of discrete codecs.

## DATASET DESCRIPTION AND LINGUISTIC ANALYSIS

This study utilizes the Idayu subset of the open-source Malaysian-TTS dataset released by the Mesolitica research team [15]. The corpus comprises high-fidelity recordings from a single female native Malay speaker, totalling 12,421 utterances and yielding approximately 12.4 hours of continuous speech. All acoustic waveforms were recorded in a controlled studio environment and strictly standardized to a 24 kHz sampling rate with 16-bit PCM encoding.

We investigate the normalized letter distributions of the speech corpus across the 12,421 training utterances for statistical analysis. Prior to counting the characters, all punctuation marks and symbols are removed from the transcripts, and uppercase letters are converted to lowercase so that the sequences contain only alphabetical letters. Figure 3 shows the resulting relative frequency distribution of the 26 letters across the dataset. The data demonstrates that the letter 'a' exhibits the highest occurrence frequency at 17.0 percent. This is followed by the consonant 'n' at 10.0 percent, along with the vowels 'e' at 8.5 percent and 'i' at 7.7 percent.

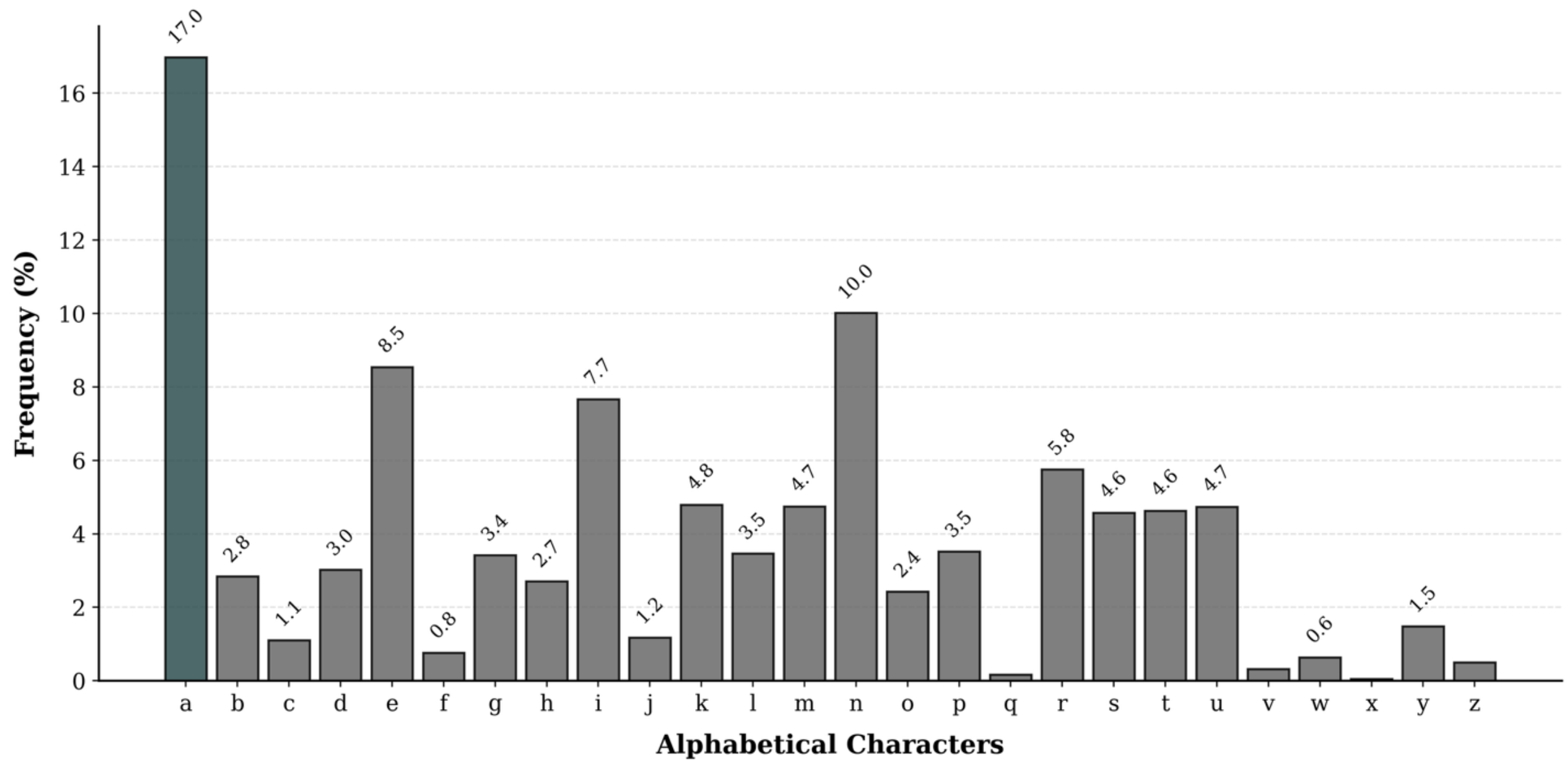

*Figure 3.* Letter Frequency Distribution in the Idayu Corpus

The distribution shows that letter frequencies vary across the corpus, with common letters occurring more frequently than others. Letters such as ‘q’, ‘v’, and ‘x’ appear less frequently than the more common letters in Malay text. This variation provides an overview of the character-level input used for forced alignment and model training. However, letter frequency alone does not indicate character duration or acoustic realization, which must be examined separately.

## METHODOLOGY: DATA PROCESSING PIPELINE AND ACOUSTIC MODEL INTERGRATION

This section presents a two-phase framework for lexicon-free Malay speech synthesis using discrete neural audio codec tokens. The framework combines a dual-track data preparation pipeline with a dual-resolution synthesis architecture based on FastSpeech 2. The data preparation pipeline generates character-level alignment information and extracts acoustic features at two temporal resolutions for model training. As shown in Figure 4, the pipeline comprises four sequential stages: text cleaning, character-level forced alignment, dual-track acoustic feature extraction, and global statistical compilation. The second phase presents DRC-TTS, the proposed dual-resolution codec-based synthesis architecture built upon FastSpeech 2. The model uses a temporal resolution of 50 Hz for character-level length regulation and subsequently downsamples the representations to 12.5 Hz for discrete codec-token prediction. By separating the temporal resolution used for length regulation from the frame rate used for codec-token generation, the architecture aims to reduce zero-frame truncation while maintaining compatibility with the codec's frame rate.

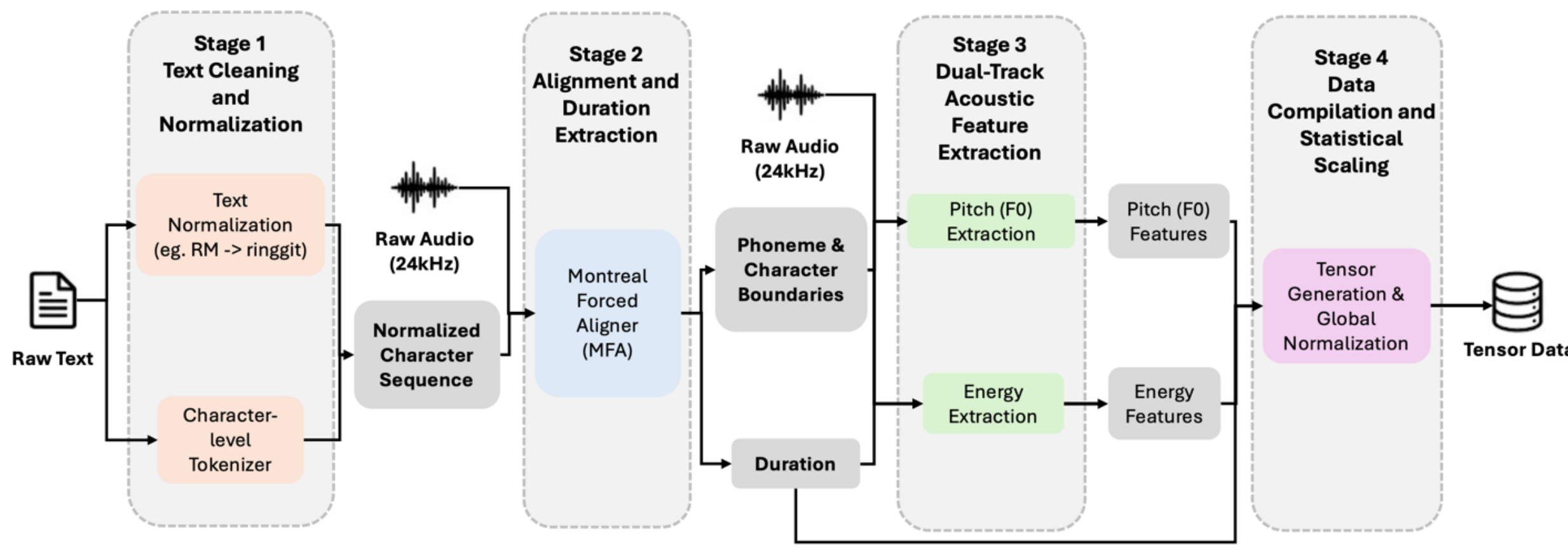

*Figure 4.* The Data Preparation Pipeline for Malay Discrete Speech Synthesis

## Text Cleaning and Normalization

The initial text processing phase standardizes the raw corpus transcripts to ensure accurate character to acoustic alignment. All alphabetical characters are converted to lowercase, and non-phonetic punctuation marks are systematically removed using regular expression filtering. This cleaning process reduces acoustic noise during training and provides a consistent textual sequence for the sequence encoder.

Following the initial formatting, a domain specific normalization algorithm processes complex semantic classes. Numerical digits, localized currency symbols, and compound date formats are transcribed into their corresponding Malay orthographic words. This expansion guarantees that the textual sequence length directly matches the spoken acoustic output, which is strictly required for lexicon free synthesis. Table 2 outlines specific examples of the numerical and symbolic expansions applied across the dataset. Finally, utterances containing irregular graphemes that cannot be mapped through standard Malay orthography are excluded from the training distribution.

Table 2

*Examples of Malay Numeric and Symbol Expansion Rules*

| Rule Type | Original Text | Expanded Text |
|---|---|---|
| **Currency** | RM 10.50 | sepuluh ringgit lima puluh sen |
| **Year / Date** | 2026 | dua ribu dua puluh enam |
| **Decimals** | 3.14 | tiga perpuluhan satu empat |
| **Ordinal Numbers** | ke-5 | kelima |

## Alignment and Duration Extraction

Because the Malay language features a highly regular spelling system, the proposed pipeline implements a character level tokenization strategy. The normalized text is tokenized into individual character units. To align these characters with the corresponding speech audio, the Montreal Forced Aligner [26] is employed alongside a pre trained acoustic model. This alignment phase is essential to extract the target duration values required to train the duration predictor within the acoustic model.

As illustrated by the Praat [36] visualization in Figure 5, non-linguistic acoustic events, such as silence intervals, are mapped to a special token rather than arbitrarily stretching the adjacent characters. The absolute start and end times for both the characters and the silence tokens are extracted from the aligner outputs. The physical duration of each character is subsequently calculated from these timestamps and converted into frame level durations.

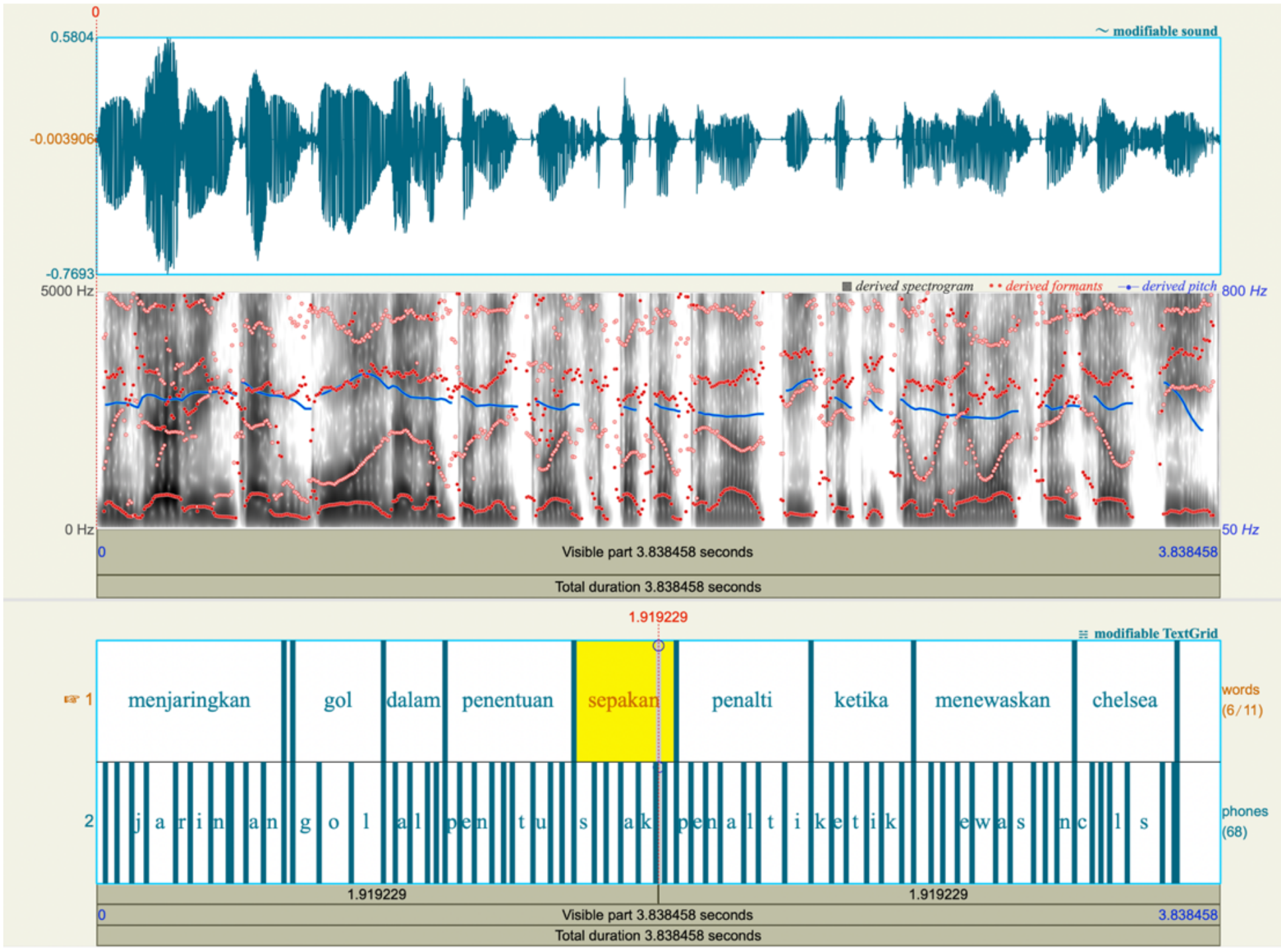

*Figure 5.* Visualization of MFA Alignment on a Malay Utterance

**Dual-Track Acoustic Feature Extraction**

Although the sequence encoder captures relationships between adjacent characters, the length regulator must still assign a discrete frame count to each individual character token. Integrating this character level tokenization directly with low framerate discrete codecs introduces a temporal alignment limitation. The external discrete audio codec used in this architecture is the Mimi codec, configured to operate at a base frame rate of 12.5 Hz using 32 parallel quantizers. At this frame rate, one acoustic frame represents 80 milliseconds of speech. Because individual characters frequently exhibit physical durations shorter than 80 milliseconds, standard duration rounding forces rapid tokens to receive zero frames. Specifically, when char [27] acter duration estimates are rounded to frame counts at 12.5 Hz, physical durations below 40 milliseconds round down to zero. For instance, a short consonant such as the character 't' undergoes this duration rounding, causing the physical erasure of the token from the training data. Consequently, these erased tokens contribute no expanded hidden states during length regulation, inherently leading to pronunciation omissions.

To address this structural bottleneck without changing the efficiency of the discrete codec, the internal temporal processing of the acoustic model is hierarchically separated. A dual-track feature extraction methodology processes acoustic features along two distinct temporal pathways associated with different temporal resolutions. The character duration sequences are extracted at a temporal resolution of 50 Hz, where each frame represents 20 milliseconds. During this extraction, the calculated character durations are mathematically rounded to the nearest integer. To prevent zero frame assignments, the implementation enforces a minimum duration constraint of one frame for every character token. Assigning a rapid consonant to occupy at least one 20 millisecond frame ensures its linguistic hidden state is preserved within the network structure.

Concurrently, the continuous prosodic features are extracted and aligned to the 12.5 Hz codec frame rate. The fundamental frequency is computed using the YIN algorithm [37]. To reflect the physiological vocal range of the selected speaker, the extraction parameters are bounded between 50 Hz and 500 Hz. To approximate human auditory perception and enhance the numerical stability of the neural network during optimization, the raw fundamental frequency trajectories are transformed into the logarithmic domain via a mathematical log operation. The signal energy is quantified using a frame level root mean square calculation. To ensure temporal synchronization across the dual track system, the higher-resolution prosodic contours are segmented based on the 12.5 Hz frame boundaries. A slice based averaging operation computes the mean fundamental frequency and energy for each segment. This segmentation ensures the prosodic sequence lengths precisely match the 12.5 Hz discrete token targets, providing a stable foundation for statistical scaling.

**Data Compilation and Statistical Scaling**

The processed linguistic and acoustic features are compiled to facilitate batch training. During dataset compilation, utterances of varying lengths are padded to form uniform tensors. Silent frames mapping to zero values are excluded to prevent statistical skewing. The system aggregates the global mean, standard deviation, minimum, and maximum values for both the fundamental frequency and energy contours. These derived metrics initialize the standard scaler within the acoustic model. The continuous input features are centered and scaled prior to network optimization. Table 3 presents the statistics of the processed corpus. The resulting training inputs comprise normalized character sequences, character durations represented at 50 Hz, alongside prosodic features and codec targets at 12.5 Hz.

Table 3

*Final Statistics of the Processed Idayu Corpus*

| Corpus Metric | Quantitative Value |
|---|---|
| **Total Utterances** | 12,421 |
| **Total Recording Duration** | ~12.4 hours |
| **Unique Vocabulary Size** | 9,752 |
| **Total Character Tokens** | 745,436 |
| **Average Character Duration** | 57.04 ms |

## Dual-Resolution Acoustic Model Integration

The extracted features are integrated into a dual resolution non autoregressive acoustic architecture based on the FastSpeech 2 framework. This structural design addresses the zero frame truncation problem by separating the temporal expansion resolution from the discrete decoding resolution. As illustrated in Figure 6, the model expands character representations at 50 Hz and subsequently downsamples the sequence to 12.5 Hz for acoustic token prediction.

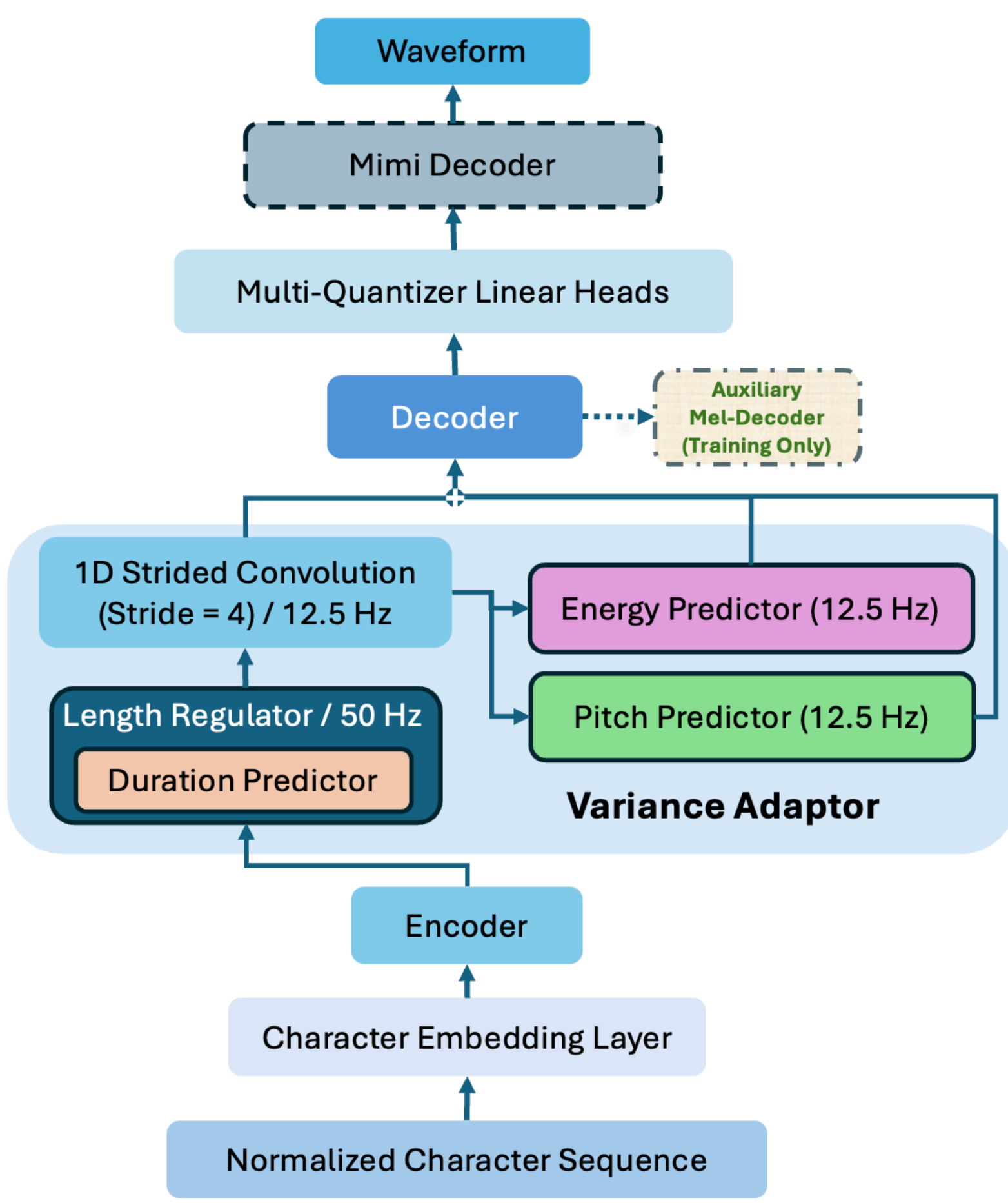

*Figure 6.* Dual-resolution Architecture

The architecture comprises a character embedding layer, a sequence encoder, a hierarchical variance adaptor, a strided convolutional downsampling layer, and a discrete token decoder. First, the normalized character sequence is projected into a continuous vector space and processed by the encoder to obtain linguistic representations. Within DRC-TTS, the dual-resolution design separates character-level length regulation from codec-token prediction. To preserve the representations of short-duration characters, the length regulator operates at a temporal resolution of 50 Hz. The encoder outputs are expanded according to the corresponding 50 Hz duration targets, ensuring that each character receives at least one frame in the expanded hidden sequence.

Following length regulation, the architecture bridges the temporal resolution gap between the 50 Hz hidden representations and the 12.5 Hz codec frame rate. The expanded 50 Hz hidden states are processed by a one-dimensional convolutional layer with a stride of four, reducing the temporal sequence to 12.5 Hz. Although the length regulator assigns at least one frame to each character at 50 Hz, the corresponding hidden representations may contribute to the downsampled sequence through the convolutional layer's local receptive field. This operation allows information from short-duration characters to contribute to the 12.5 Hz representations rather than being necessarily discarded during length regulation.

Once downsampled to 12.5 Hz, the hidden states are combined with the prosodic features. The pitch and energy predictors are trained against the corresponding 12.5 Hz targets extracted during preprocessing. Aligning the prosodic features with the codec frame rate maintains temporal consistency and avoids introducing a higher-rate prosodic sequence into the lower-rate codec-token prediction stage.

Following the FastSpeech 2 framework, the sequence enriched with prosodic information is processed by the decoder's feed-forward Transformer blocks. The decoder produces a sequence of hidden-state vectors at 12.5 Hz, with each vector corresponding to one codec frame. To interface with the discrete audio codec, the architecture uses 32 linear projection heads, one for each quantizer level of the Mimi codec. Token prediction is parallel across time but sequential across the quantizer levels for each codec frame.

## Training Objectives and Inference Mechanisms

During the training phase, the network jointly minimizes the duration, pitch, energy, and discrete token prediction losses. To guide the intermediate representations, an auxiliary continuous Mel-spectrogram loss is incorporated into the training objective. As reflected by the structural branches in the proposed architecture, the decoder hidden states project to continuous Mel-spectrograms alongside the discrete tokens. This auxiliary mechanism functions as an internal regularizer to stabilize gradient convergence during the initial training steps.

During inference, the model operates without an explicit pronunciation lexicon. As illustrated by the forward pass in Figure 6, the architecture predicts character durations using the trained 50 Hz duration predictor. These predicted durations are mathematically rounded and constrained to a minimum of one frame to drive the length regulator. The expanded 50 Hz sequence is subsequently downsampled to 12.5 Hz via the strided convolution. This downsampled sequence is then enriched with the predicted pitch and energy features and processed by the decoder. Finally, the sequentially predicted multi-level discrete tokens are fed directly into the external Mimi codec decoder module to synthesize the acoustic waveform.

## VALIDATION AND EXPERIMENTAL RESULTS

To evaluate the proposed linguistic processing pipeline and DRC-TTS, a comprehensive evaluation framework was established. This section describes the experimental setup used to train the acoustic models. An objective analysis of character-level quantization is then presented to investigate the effect of the proposed dual-resolution design. Subjective perceptual evaluations assess the naturalness of the synthesized speech against a single-resolution baseline. Finally, qualitative stress tests using complex semantic structures are conducted to examine the robustness of the text normalization pipeline.

### Experiment Setup

To ensure reproducibility, all validation procedures were conducted within a consistent hardware and software environment. Model training was executed on a single NVIDIA GeForce RTX 4090 graphics processing unit equipped with 24 gigabytes of video random access memory using the PyTorch deep learning framework. Prior to model training, the processed Idayu speech corpus was partitioned into designated subsets to prevent data leakage. The dataset was divided to allocate ninety percent of the utterances for network training, five percent for intermediate validation, and the remaining five percent for unseen testing and evaluation.

Two non-autoregressive neural systems were trained for comparison. The baseline uses a single-resolution architecture operating entirely at 12.5 Hz, following the frame rate of the discrete audio codec. The proposed DRC-TTS model uses a dual-resolution architecture, with character representations expanded at 50 Hz before being downsampled to 12.5 Hz for discrete codec-token prediction. This comparison evaluates the effect of introducing a higher temporal resolution for character-level length regulation while retaining the codec's original frame rate.

During the optimization phase, both networks were trained to minimize the duration, pitch, energy, and discrete token prediction losses jointly. A continuous Mel spectrogram auxiliary loss was incorporated into the training objective to guide the intermediate hidden states. The AdamW optimizer was employed alongside a Noam learning rate scheduling strategy incorporating an initial warm up phase. The training process spanned 150,000 steps utilizing a batch size of 4, ensuring network convergence across both the baseline and proposed architectures.

**Objective Analysis: Character Quantization**

To quantify the resolution gap, an objective analysis focused on the mathematical distribution of raw character durations prior to temporal quantization. The ground truth character durations were analyzed directly from the forced alignment labels across the corpus. In a conventional single resolution architecture operating at 12.5 Hz, each acoustic frame represents 80 milliseconds. According to standard quantization logic, characters with a physical duration shorter than 40 milliseconds are rounded to zero frames.

Statistical analysis of the dataset revealed that 20.23% of the character tokens have estimated durations shorter than 40 ms. The proposed DRC-TTS model addresses the resulting temporal alignment challenge through its dual-resolution design. Character durations are represented at a temporal resolution of 50 Hz, allowing the length regulator to allocate at least one 20 ms frame to each character. This minimum-duration constraint prevents characters from receiving zero frames during high-resolution length regulation. In contrast, a single-resolution baseline operating directly at 12.5 Hz may assign zero frames to sufficiently short-duration characters, depending on the duration-quantization procedure. By preserving character representations during the high-resolution expansion stage, DRC-TTS provides a more complete hidden sequence for subsequent downsampling and discrete codec-token prediction.

Statistical analysis of the dataset determined the exact frequency of these extremely short linguistic tokens. As illustrated by the character duration distribution histogram in Figure 7, a significant portion of the consonant characters falls within the sub threshold erasure zone. Specifically, the data revealed that 20.23 percent of the character tokens possess raw physical durations under 40 milliseconds. This establishes a baseline zero frame quantization rate of 20.23 percent.

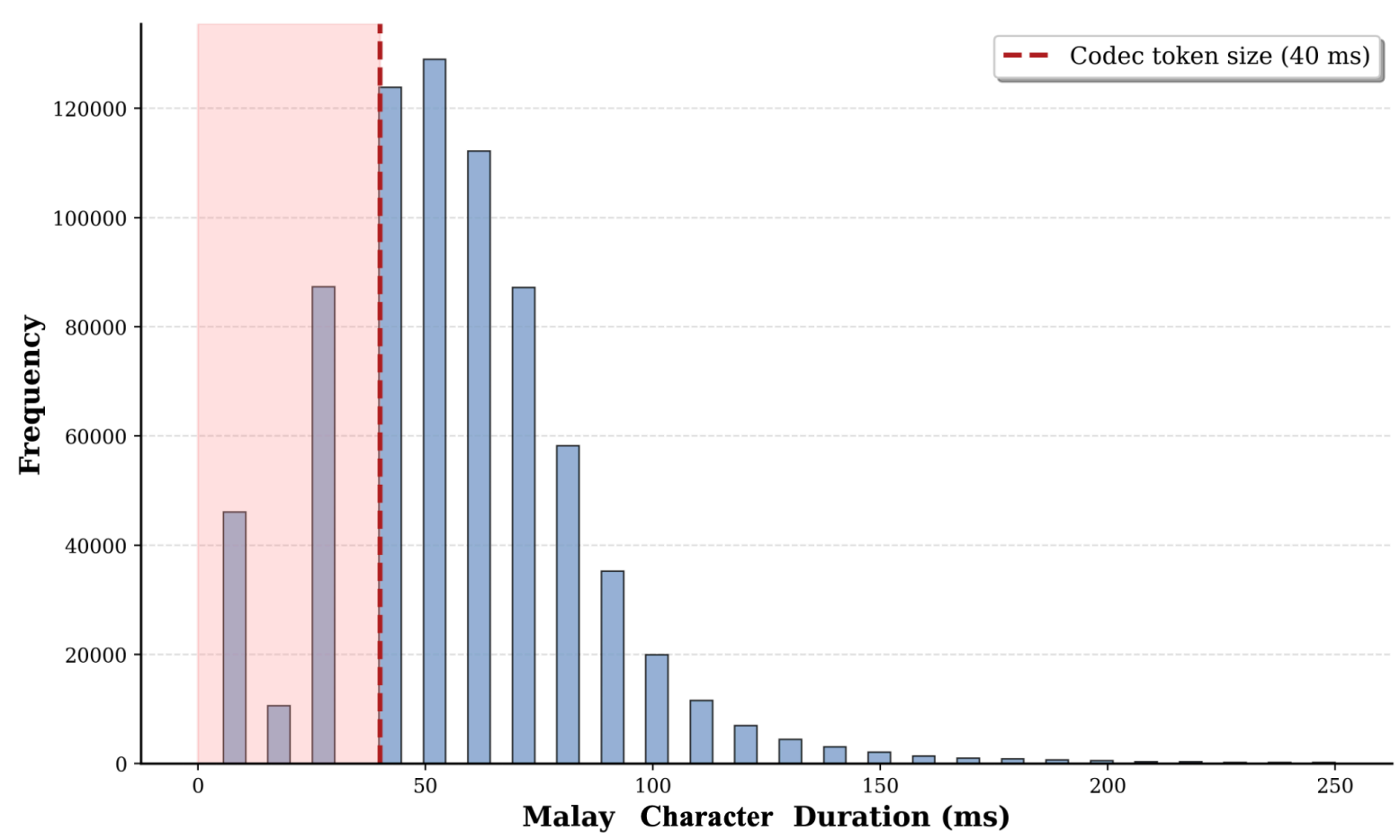

*Figure 7.* Duration distribution in the Malay Idayu corpus

The proposed dual-resolution architecture addresses this temporal alignment challenge by representing character durations at a temporal resolution of 50 Hz and operating the length regulator at the same frame rate. This allows the architecture to allocate a 20 ms frame to each character. The model also enforces a minimum duration constraint of one frame per character, preventing zero-frame assignments during high-resolution length regulation. By preserving character representations at this stage, DRC-TTS provides a more complete hidden sequence for subsequent downsampling and discrete codec-token prediction.

### Objective Analysis: Acoustic Evaluation

To quantitatively assess the acoustic fidelity and prosodic accuracy of the synthesized speech, a comprehensive objective evaluation was conducted. The synthesized audio samples from both the baseline and the proposed architectures were compared against the ground truth recordings using four standard acoustic metrics: Mel-Cepstral Distortion (MCD), Band Aperiodicity (BAP), Fundamental Frequency Root Mean Square Error (F0 RMSE), and Voiced/Unvoiced (V/UV) error rate. MCD and BAP measure the spectral distance and acoustic detail preservation, while F0 RMSE and V/UV error evaluate the accuracy of the synthesized prosody and pitch contours.

The objective acoustic evaluation results are presented in Table 4. The single-resolution baseline, which operates at 12.5 Hz, achieved an MCD of 11.60 dB and an F0 RMSE of 41.59 Hz. In comparison, the proposed DRC-TTS model, which uses 50 Hz temporal expansion before downsampling to the 12.5 Hz codec frame rate, achieved an MCD of 10.70 dB and an F0 RMSE of 40.32 Hz. The V/UV error rate also decreased to 2.01%, while the band aperiodicity results remained comparable to those of the baseline. These results indicate that the dual-resolution design is associated with improvements in the evaluated spectral and prosodic metrics under the experimental conditions. The findings are consistent with the hypothesis that higher-resolution character-level length regulation can help preserve short-duration linguistic representations.

Table 4

*Acoustic Fidelity Comparison on Baseline System*

| System | MCD (dB) | BAP (dB) | F0 RMSE (Hz) | V/UV ERROR (%) |
|---|---|---|---|---|
| **Baseline (Single-Resolution 12.5 Hz)** | 11.60 | 11.45 | 41.59 | 2.06 |
| **Proposed (Dual-Resolution 50 Hz/12.5Hz)** | 10.70 | 11.46 | 40.32 | 2.01 |

### Objective Analysis: Pronunciation Accuracy

To objectively measure the intelligibility and pronunciation accuracy of the synthesized speech, the Character Error Rate (CER) was computed utilizing an external pre trained automatic speech recognition system. This metric evaluates whether the acoustic models successfully generate complete phonetic sequences without internal character omissions. The quantitative results comparing the ground truth recordings with the synthesized outputs are detailed in Table 5.

Table 5

*Character Error Rate Evaluation on Synthesized Speech*

| System | Audio Source | Character Error Rate % |
|---|---|---|
| **Ground Truth** | Original Idayu Recordings | 16.32 |
| **Baseline System** | Single-Resolution 12.5 Hz | 27.11 |
| **Proposed System** | Dual-Resolution (50 Hz/12.5 Hz) | 21.86 |

As demonstrated in Table 5, the ground truth recordings established a baseline recognition error rate of 16.32 percent. This value defines the physical upper bound of the recognition performance on this specific localized Malay corpus. The baseline single resolution architecture yielded an elevated error rate of 27.11 percent. This high error percentage provides objective acoustic evidence of the pronunciation omissions occurring under the strict 12.5 Hz resolution constraint, corroborating the zero frame truncation issue identified in the previous temporal analysis.

In contrast, the proposed dual-resolution architecture achieved a CER of 21.86%, lower than the baseline's 27.11%. This result suggests that using a higher temporal resolution for character-level length regulation can help reduce character recognition errors in the synthesized speech. The improvement is consistent with the motivation for reducing zero-frame truncation during duration quantization, although the CER results alone do not directly establish the extent to which this truncation is reduced.

**Subjective Analysis: Perceptual Evaluation**

To assess the naturalness, intelligibility, and overall audio fidelity of the synthesized speech, a Mean Opinion Score (MOS) evaluation was conducted. The subjective evaluation panel comprised twenty native Malay speakers to ensure accurate judgment of localized prosody and pronunciation patterns. A set of twenty out of domain sentences, randomly sampled from the withheld testing distribution, was utilized to evaluate the original recordings, the baseline single resolution system, and the proposed DRC-TTS model. The evaluators listened to the audio samples using professional audio equipment in an isolated environment and rated the naturalness on a standard five point scale. The quantitative results accompanied by their 95 percent confidence intervals are detailed in Table 6.

As demonstrated in Table 6, the baseline single-resolution architecture achieved a mean opinion score (MOS) of 4.12. Evaluators noted occasional occurrences of truncated words and missing consonants. These observations are consistent with the zero-frame quantization issue identified in the objective analysis. The proposed DRC-TTS model achieved a higher MOS of 4.30, indicating a descriptive improvement over the baseline. Although a larger evaluator pool and formal hypothesis testing are needed to establish statistical significance, this result suggests a potential perceptual benefit from preserving short-duration character representations through higher-resolution temporal expansion. The ground-truth recordings achieved a MOS of 4.65, indicating a remaining gap between the proposed system and natural speech. This gap may partly reflect acoustic artifacts introduced by the discrete neural codec under high compression. Overall, these subjective results suggest that DRC-TTS may improve the perceived naturalness of synthesized speech under the evaluated conditions.

Table 6

*Mean Opinion Score (MOS) Results with 95% Confidence Intervals*

| System | Audio Source | MOS ± 95% CI |
|---|---|---|
| **Ground Truth** | Original Idayu Recordings | 4.65 ± 0.08 |
| **Baseline System** | Single-Resolution 12.5 Hz | 4.12 ± 0.09 |
| **Proposed System** | Dual-Resolution (50 Hz/12.5 Hz) | 4.30 ± 0.15 |

### Qualitative Analysis: Text Normalization

To evaluate the robustness of the frontend text normalization protocols, a qualitative stress testing methodology was implemented. This stress testing evaluates the structural integrity of the linguistic processing pipeline by forcing the acoustic model to synthesize complex semantic edge cases. The test set included mixed decimal currencies, compound date formats, ordinal numbers, and localized abbreviations. In conventional lexicon based systems, these structures frequently trigger out of vocabulary errors, leading to pronunciation omissions or alignment failures.

By exposing the model to these challenging inputs, the evaluation assesses the effectiveness of the text cleaning and domain specific normalization stages outlined in the methodology. The qualitative results are detailed in Table 7. The input text represents the raw string provided to the inference engine, while the synthesized pronunciation reflects the explicit phonetic sequence generated by the acoustic model.

Table 7

*Qualitative Evaluation on Complex Semantic Structures*

| Sentence Type | Input Text | Actual Synthesized Pronunciation | Pronunciation Accuracy |
|---|---|---|---|
| **Decimal Currency** | Harga telefon ini ialah RM 3,450.99. | harga telefon ini ialah tiga ribu empat ratus lima puluh ringgit sembilan puluh sembilan sen | Successfully converted into valid words. |
| **Compound Date** | Projek ini bermula pada 15 Ogos 2026. | projek ini bermula pada lima belas ogos dua ribu dua puluh enam | Symbols have been successfully converted into valid words with a natural intonation. |
| **Ordinal Numbers** | Beliau memenangi tempat ke-3 dalam acara itu. | beliau memenangi tempat ketiga dalam acara itu | Successfully converted into valid words. |
| **Time / Abbreviations** | Mesyuarat dijadualkan pada pukul 08:30 a.m. | mesyuarat dijadualkan pada pukul lapan tiga puluh pagi | Successfully converted into valid words. |

As evidenced by Table 7, the synthesized audio maintains stable and accurate pronunciation across all tested semantic categories. The model articulates lengthy numerical sequences and expands localized symbols effectively without exhibiting acoustic degradation. The synthesized pronunciation corresponds exactly to the expanded phonetic sequence. This execution provides qualitative evidence that the proposed frontend pipeline standardizes ambiguous inputs effectively. By resolving complex graphemes prior to acoustic feature mapping, the pipeline mitigates common pronunciation failures. This standardization establishes the alignment stability strictly required for lexicon free Malay speech synthesis.

## CONCLUSION

This study presents DRC-TTS, a dual-resolution codec-based architecture for lexicon-free Malay speech synthesis. The proposed approach combines character-level text processing with discrete neural audio codec tokens to address the temporal alignment challenges associated with low-frame-rate codec representations. The framework incorporates text normalization and character-level modeling to support Malay speech synthesis without requiring a pronunciation lexicon during inference.

DRC-TTS performs character-level length regulation at a temporal resolution of 50 Hz and subsequently downsamples the hidden representations to 12.5 Hz for discrete codec-token prediction. By separating the resolution used for character-level length regulation from the codec frame rate, the proposed design aims to reduce the loss of short-duration character representations during duration quantization while maintaining compatibility with the codec.

Experimental results show that DRC-TTS reduces the character error rate from 27.11% for the baseline to 21.86%, an absolute reduction of 5.25 percentage points. The proposed model also achieves an MCD of 10.70 dB, compared with 11.60 dB for the baseline, and reduces the F0 RMSE from 41.59 Hz to 40.32 Hz. In the subjective evaluation, DRC-TTS obtains a mean opinion score of 4.30, compared with 4.12 for the baseline and 4.65 for the original recordings. These findings suggest that the dual-resolution design offers potential benefits for character-level alignment and synthesized speech quality under the evaluated conditions. However, further experiments are needed to establish the generalizability of these findings and determine whether the observed perceptual differences are statistically significant.

Future work will investigate the scalability of DRC-TTS across larger and multi-speaker Malay speech datasets, as well as its applicability to other languages. Further analysis of duration quantization and temporal alignment may also help clarify the contribution of the dual-resolution design to synthesis quality.